\documentstyle[12pt]{article}

\catcode`\@=11

\begin{document}

\baselineskip 24pt
\newcommand{\numero}{SWAT/40}
\newcommand{\titre}{ADDITIONAL FERMION FAMILIES AND PRECISION}
\newcommand{\titreb}{ELECTROWEAK DATA}
\newcommand{\auteura}{Nick Evans}
\newcommand{\addressa}{ }
\newcommand{\auteurc}{D.A. Ross }
\newcommand{\beq}{\begin{equation}}
\newcommand{\eeq}{\end{equation}}
\newcommand{\Fn}{\mbox{$F(p^2,\Sigma)$}}

\newcommand{\addressc}{Department Of Physics  \\   University   of
     Wales, Swansea\\ Singleton park, \\ Swansea \\ SA2 8PP \\ U.K. }
\newcommand{\abstrait}{ The S,T, and U formalism for studying electroweak
precision data has recently been updated to include the effects of new light
physics
and within this analysis the latest LEP and SLC data tightly constrain
models of physics beyond the Standard Model.
We re-examine the constraints on additional strong and weakly interacting
fermion families in the light of these developments. We conclude that the
precision data favour models with at least one fermion with a mass below
150GeV.  }

\begin{titlepage}
\hfill \numero
\vspace{.5in}
\begin{center}
{\large{\bf \titre }}
{\large{\bf \titreb}}
\bigskip \\by\bigskip\\ \auteura \bigskip \\ \addressc \\

\renewcommand{\thefootnote}{ }
\vspace{.9 in}
{\bf Abstract}
\end{center}
\abstrait
\bigskip \\
\end{titlepage}

\def\id{\rlap{1}\hspace{0.15em}1}

The most recent precision data from LEP and SLC \cite{data} have further
tightened the constraints on  models of new physics beyond the standard
model. In addition the confusion between a number of different
parameterizations of the observable radiative correction parameters
\cite{Peskin,Barbieri,Burgess1} in the
literature has recently been clarified \cite{Burgess2}. The oblique corrections
from
heavy physics ($m_{new}>M_Z/2$) to low energy observables on the Z pole
may be parameterized by the two variables \cite{Burgess2,Barbieri}

\beq \begin{array}{ccccc}
S' & = & S + 4s_w^2c_w^2V + 4(c_w^2-s_w^2)X & = &
          0.0084(\epsilon_3-\epsilon_3^{SM}) \\
&&&&\\
T' & = & T + V & = & 0.0078(\epsilon_1-\epsilon_1^{SM}) \
\end{array} \eeq

A fit to the most recent LEP and SLC data \cite{data} performed by
Caravaglios \cite{Caravaglios} gives (with $m_{top} = 150GeV$ and
$m_{H} = 300GeV$) the one standard deviation bounds

\beq \begin{array}{c}
-0.49 < S' < -0.036 \\
\\
-0.13 < T' < 0.32
\end{array} \eeq

It is not sufficient to only consider the oblique corrections to low energy
observables \cite{Kitazawa,Barbieri}. Large non-oblique corrections to the
$Zb{\bar b}$
vertex may affect the magnitude of $S'$ and $T'$.  This additional constraint
may be measured by the parameter \cite{Barbieri}

\beq B = 2.0S' - 4.8T' +436 \frac{\delta \Gamma_b}{\Gamma_0} = (\epsilon_b
      -\epsilon_b^{SM}) \times 10^{3} \eeq

\noindent where $\Gamma_0 = 379.4MeV$.
The experimental fit gives at one standard deviation
(with $m_{top} = 150GeV$ and $m_{H} = 300GeV$)

\beq 1.58 < B < 9.98 \eeq

The bounds on $S'$ and $T'$ in Eqn(2) correspond to a contribution
$-2.5<B<0.6$ and thus the deviation from zero in Eqn(4) is not
compatible with $\delta \Gamma_b = 0$. Provided $0.94\% < \delta\Gamma_b /
\Gamma_0
< 2.2\%$ the B parameter provides no additional constraint on $S'$ and
$T'$ beyond Eqn(2). If $\delta\Gamma_b / \Gamma_0$
rises to 2.9$\%$ or falls to $0.2\%$ then Eqn(4) over constrains $S'$ and $T'$
and there is no compatible solution (to one standard deviation).
Since the B parameter bounds are dominated by
the uncertainty in the non oblique correction $\delta \Gamma_b$ we shall
henceforth make the approximation that  Eqn(4) places no additional constraint
beyond Eqn(2) on the oblique parameters $S'$ and $T'$. We note that models
including additional gauge bosons with masses of a few TeV can give rise to
both positive and negative contributions to B \cite{negB,0B,posB} though we
shall concentrate
on the oblique correction bounds on additional fermions below.

In the light of these new results it is interesting to re-examine
the constraints on additional strong and weakly interacting fermion families.

\noindent {\bf A Weakly Interacting Fourth Family}

We consider a fourth weakly interacting family (U,D,N,E) with Dirac masses
in addition to the
usual fermion content of the Standard Model. In Figs 1 and 2 we show the
contribution
of the fourth family to the oblique parameters in the $S' T'$ plane. The points
are generated by allowing each of the fermions' masses to vary in 50GeV steps
between 50 GeV (the LEP lower bound) and 500GeV and span the $S' T' $
parameter space of fourth family models. Fig 1 shows the numerical results
when all the fermions have masses $\geq 150GeV$. Fig 2 shows results
for when at least one family member has a mass $\leq 100GeV$.
The ellipses mark the one and two standard
deviation experimental bounds. The experimental data is not incompatible with a
fourth family but favours models with at least one light  fermion ($m_f  <
150GeV$).

\noindent {\bf A Technifamily}

An additional strongly interacting ``technifamily" \cite{ETC} will have
$N_{TC}$
degenerate families depending upon the
strong interaction group. We shall take $N_{TC} = 3$ to demonstrate the
experimental constraints. The most naive estimate of such an additional
family's
contribution to $S'$ and $T'$ is to treat the fermions as weakly interacting
and
is shown in Fig 3. In this approximation the experimental ellispes ly entirely
within the parameter space available to such a model. The trend towards light
physics is, however, enhanced relative to the single family case discussed
above
(and would be more so were $N_{TC}$ higher). We note that the technifamily
spectrum

\beq MU = MD=300GeV, \hspace{0.2cm} MN = 50GeV,  \hspace{0.2cm} ME = 150GeV
\eeq

\noindent proposed in Ref\cite{Revenge} is still consistent in this
approximation
and with the new data. Ref\cite{NJE} has demonstrated that extended
technicolour models
can be built consistent with the third family masses and giving rise to this
spectrum.

Including the resonance effects of the strong interactions is a
non-perturbative
calculation, however, a number of approximations have been made in the
literature
\cite{Peskin,NLCM,NPS}. In the isospin preserving limit a good estimate can be
made by scaling
low energy QCD results \cite{Peskin} or in the non local chiral model of
Ref\cite{NLCM}.
These methods are in general agreement and give

\beq \begin{array}{ccc}
S'_{\chi {\rm limit}} & \sim & 0.1 / {\rm doublet} \\
&&\\
T'_{\chi {\rm limit}} & \sim & 0 \end{array}
\eeq

\noindent The non local chiral model result for the $T'$ parameter has been
extended
into the regime where isospin is broken \cite{rho}. The results are dependent
on the
form of the fermion's dynamical self energy and hence on the precise theory of
the
strong interactions. Nevertheless the results ly in the range

\beq T'_{pert} < T' < 2T'_{pert} \eeq

\noindent and we shall use these bounds to estimate $T'$ below.

No reliable estimates for the $S'$ parameter exist away form the isospin
preserving limit
but a number of extrapolations from that limit have been made
based on the behaviour of the pertubative weak coupling results
\cite{Revenge,Maj}.
We shall estimate $S'$ per doublet by the extreme values of these estimates

\beq 2S'_{pert} < S' < S'_{pert} + 0.05 \eeq

In addition the family's strong interactions will give rise to light pseudo
Goldstone
boson bound states. In a realistic model these must have masses in excess of
the LEP
bound ($M_{PGB}>43GeV$). Their contributions to S,T,U,V,W and X have been
calculated
before \cite{Revenge,PGB} for a realistic range of masses. The results of these
references are
summerized by

\beq \begin{array}{ccccc}
0.14 & < & S' & <  & 0.27\\
&&&&\\
-0.42 & < & T' &< & 0.26 \end{array} \eeq

In Fig 4 we display the estimates and theoretical errors for the contributions
to $S'$
and $T'$ for three different mass spectra. When the fermions are all heavy and
degenerate the QCD scaled up solution is expected to be reliable but such a
spectrum is ruled out by the experimental data. Away from the isospin
preserving limit
the theoretical errors are as large as the experimental errors, however, we can
draw
some broad conclusions. Heavy non-degenerate doublets
give rise to large contributions to $T'$ and are also ruled out by the
experimental data.
Spectra such as that proposed in Refs\cite{Revenge,NJE} with light
non-degenerate
lepton doublets are, however, plausibly consistent with the experimental data.

Strongly interacting models of the Higgs sector typically give rise to
additional
heavy gauge bosons. We note that models have been proposed which give
negative \cite{negB}, zero \cite{0B} and positive \cite{posB} contributions to
$\delta \Gamma_B$ and hence the B parameter.

We conclude that recent experimental results are compatible with additional
strong or weakly interacting fermion families only when those families contain
one or more light fermion with mass less than 150 GeV.

\newpage

\noindent {\Large \bf Acknowledgements}

The author would like to thank Francesco Caravaglios for providing the results
for
the fit to the recent experimental results and SERC for supporting this work.

$\left. \right.$ \vspace{2cm}

\noindent {\bf Figure Captions}

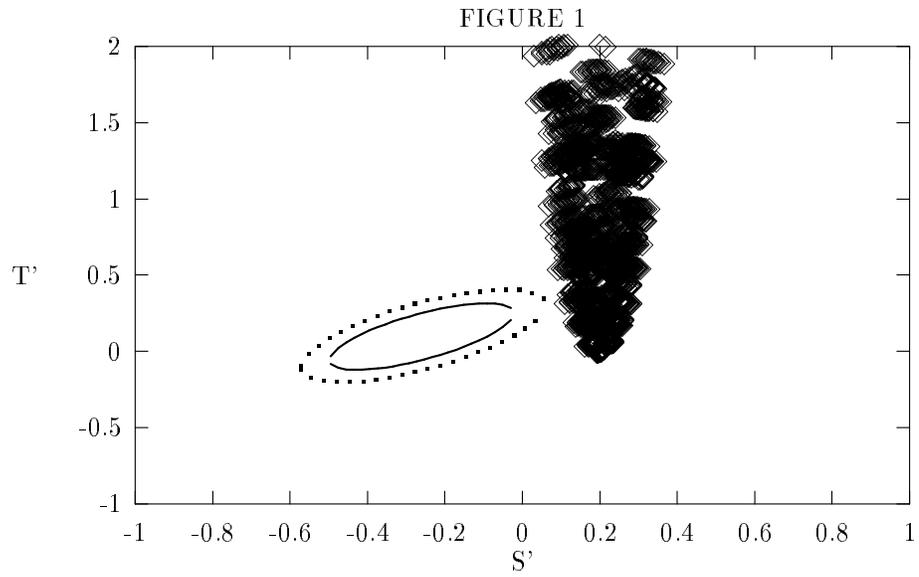
\noindent Figure 1: The contribution of a fourth weakly interacting family with
all fermion masses $\geq 150GeV$ to $S'$ and $T'$.

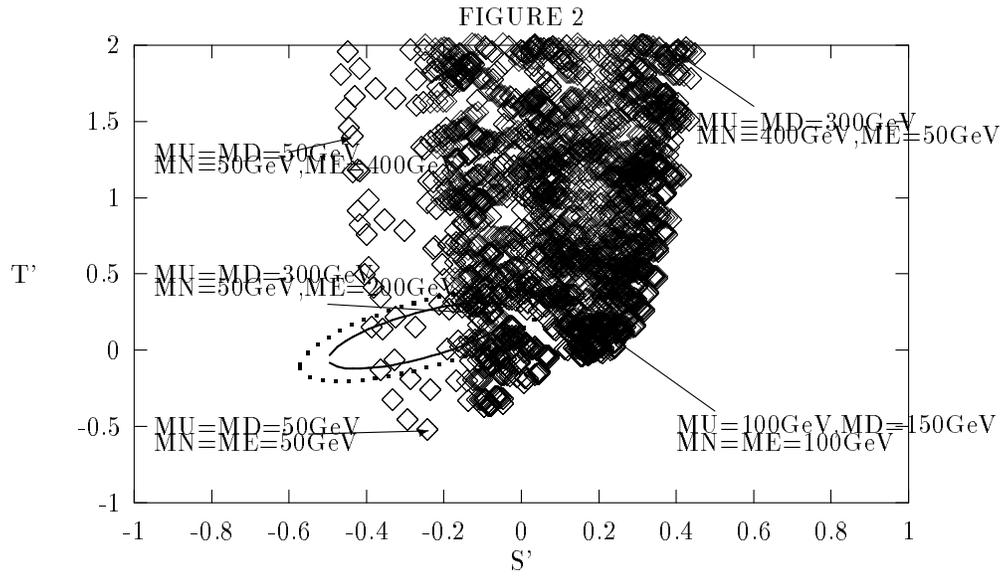
\noindent Figure 2: The contribution of a fourth weakly interacting family with
one or more fermion with mass $\leq 100GeV$.

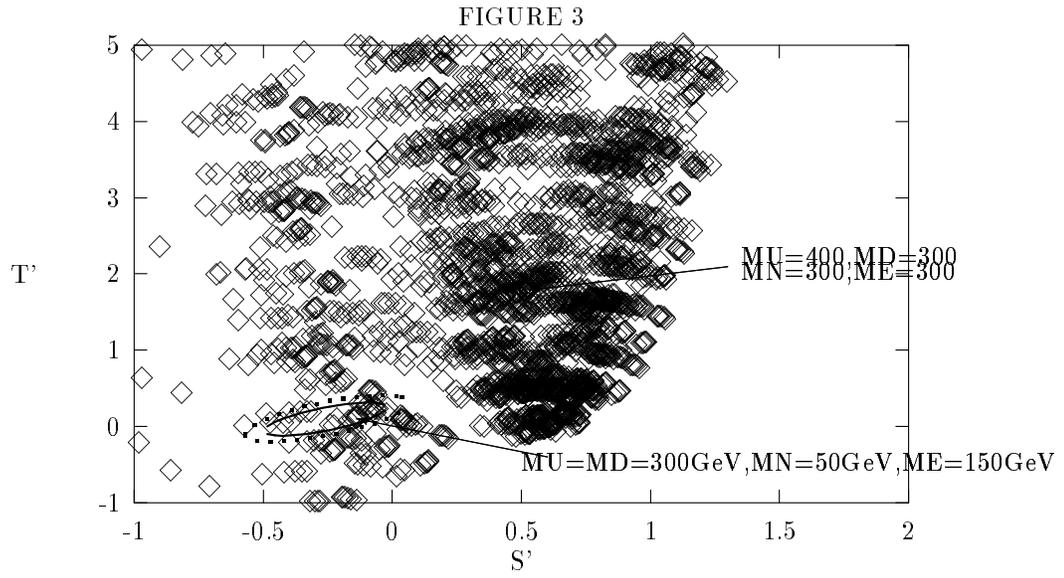
\noindent Figure 3: Weak coupling approximation to the contribution to $S'$
and $T'$ of a technifamily with $N_{TC} = 3$.

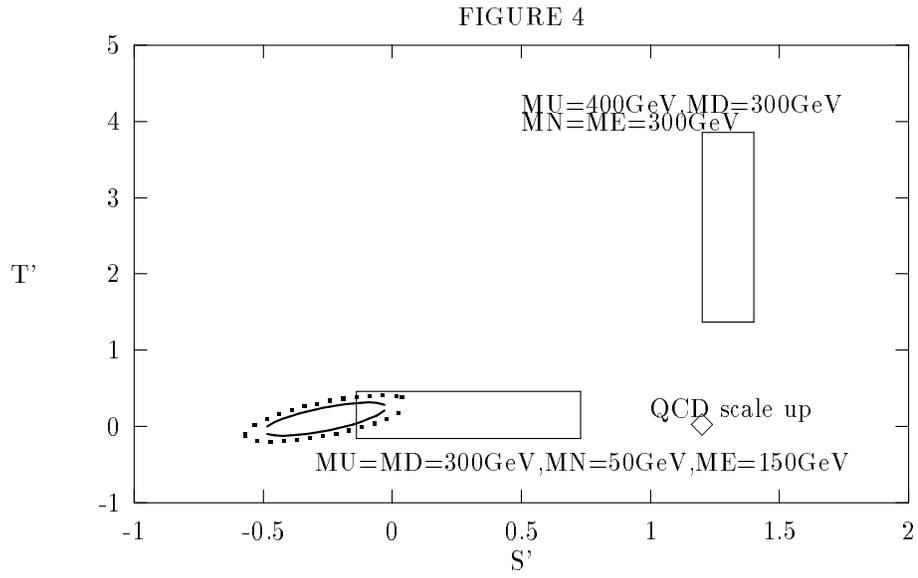
\noindent Figure 4: Estimation of contribution to $S'$ and $T'$ and theoretical
error bars for three mass spectra of strongly interacting technifamily.

\newpage

\newpage

\begin{thebibliography}{99}
 \bibitem{data} LEP and SLC presentations at Moriond '94.
 \bibitem{Peskin}M.E. Peskin and T. Takeuchi, Phys. Rev. Lett. {\bf65} (1990)
964;
      M.E. Peskin and T. Takeuchi,  Phys.  Rev. {\bf  D46} (1992) 381.
 \bibitem{Barbieri}G. Alterelli, R. Barbieri and F. Caravaglios, Nucl. Phys.
{\bf B405}
                  (1993) 3.
 \bibitem{Burgess1}C.P. Burgess, S. Godfrey, H. Konig, D. London and I.
Maksymyk,
      Phys. Lett. {\bf B326} (1994) 276; I. Maksymyk, C.P. Burgess and D.
London, McGill Preprint 93/13, to be
              published in Phys Rev D, hep-ph 9306267.
 \bibitem{Burgess2} P. Bamert and C.P. Burgess, McGill preprint, McGill-94/27
              hep-ph 9407203.
 \bibitem{Caravaglios} Private communication with Dr F. Caravaglios.
 \bibitem{Kitazawa} N. Kitazawa, Phys. Lett. {\bf B313} (1993) 395.
 \bibitem{negB}R.S. Chivukula, S.B. Selipsky and E.H. Simmons,  Phys. Rev.
Lett.
         {\bf 69} (1992) 575;R.S. Chivukula, E. Gates, E.H. Simmons and J.
Terning, Phys. Lett.
                           {\bf B311} (1993) 157.
 \bibitem{0B} N.J. Evans,  Phys. Lett. {\bf B331} (1994) 378.
 \bibitem{posB}R.S. Chivukula, E.H. Simmons and J. Terning,
                            Phys. Lett. {\bf B331} (1994) 383; B. Holdom,
University of Toronto
                           preprint, UTPT-94-20, hep-ph 9407311.
 \bibitem{ETC} E. Farhi and L. Susskind, Phys. Report 74 No.3 (1981) 277; S.
Dimopolous
                           and L. Susskind, Nucl. Phys. {\bf B155} (1979) 1277;
                           E. Eichten and K. Lane, Phys. Lett. {\bf B90} (1980)
125.
 \bibitem{Revenge}T. Appelquist and J. Terning, Phys. Lett. {\bf B315} (1993)
139.
 \bibitem{NJE} N.J. Evans, Swansea University preprint, SWAT/33 hep-ph 9405356.
 \bibitem{NPS}J. Terning, Phys. Rev. {\bf D44} (1991) 887; M. Golden and L.
Randall,
         Nucl. Phys. {\bf B361} (1991) 3; T. Yoshikawa, H. Takata and  T.
Norozumi,
         Hiroshima preprint HUPD 94-06; R. Sundrum, Nucl. Phys. {\bf B391}
(1993) 127.
 \bibitem{NLCM}B. Holdom, J.  Terning  and  K.  Verbeek, Phys. Lett.
      {\bf B245} (1990) 612;
      J. Terning, Phys Rev {\bf D44} (1991) 887;B. Holdom,
      Phys. Rev. {\bf D45} (1992) 2534.
 \bibitem{rho}  B. Holdom, Phys. Lett. {\bf B226} (1989) 137.
 \bibitem{Maj} N.J. Evans, Nucl. Phys. {\bf B417} (1994) 151.
 \bibitem{PGB} N.J. Evans, Phys. Rev. {\bf D49} (1994) 4785.
\end{thebibliography}
\end{document}